\documentclass[aps,pra,showpacs,twocolumn,superscriptaddress]{revtex4}

\usepackage{graphicx}
\usepackage{dcolumn}
\usepackage{bm}
\usepackage{amsfonts}
\usepackage{amsmath}
\usepackage{color}

\newcommand{\ket}[1]{\left\vert{#1}\right\rangle}
\newcommand{\beq}{\begin{equation}}
\newcommand{\eeq}{\end{equation}}
\newcommand{\bea}[1]{\begin{equation}\begin{array}{#1}}
\newcommand{\eea}{\end{array}\end{equation}}
\newcommand{\beqn}{\begin{eqnarray}}
\newcommand{\eeqn}{\end{eqnarray}}

\begin{document}

\title{A Family of Continuous Variable Entanglement Criteria using General Entropy Functions}

\author{A.~Saboia}
\email{saboia@if.ufrj.br}
\affiliation{Instituto de F\'{\i}sica, Universidade Federal do Rio
de Janeiro, Caixa Postal 68528, Rio de Janeiro, RJ 21941-972,
Brazil}

\author{F.~Toscano}
\affiliation{Instituto de F\'{\i}sica, Universidade Federal do Rio de
Janeiro, Caixa Postal 68528, Rio de Janeiro, RJ 21941-972, Brazil}

\author{S. P. ~Walborn}
\affiliation{Instituto de F\'{\i}sica, Universidade Federal do Rio de
Janeiro, Caixa Postal 68528, Rio de Janeiro, RJ 21941-972, Brazil}

\begin{abstract}
We derive a family of entanglement criteria for continuous variable systems based on the R\'enyi entropy of complementary distributions.  We show that these entanglement witnesses can be more sensitive than those based on second-order moments, as well as previous tests involving the Shannon entropy [Phys. Rev. Lett. \textbf{103}, 160505 (2009)].  We extend our results to include the case of discrete sampling, and develop another set of entanglement tests using the discrete Tsallis entropy.  We provide several numerical results which show that our criteria can be used to identify entanglement in a number of experimentally relevant quantum states.   
\end{abstract}


\maketitle

\section{INTRODUCTION}
\label{sec:introduction}

Quantum entanglement is a fundamental property of quantum systems that can be exploited for quantum computation, quantum teleportation and quantum cryptography \cite{chuang00}. As such,  its detection is an essential task in an experimental setting. Many techniques exist for detecting entanglement in discrete systems (see \cite{guhne09, horodecki09} for review).  In continuous variable systems,  its identification can be more complicated, due to the large Hilbert space structure. However, there is a considerable amount of work concerning entanglement detection and characterization of Gaussian states  \cite{adesso07,braunstein05} where tests involving only the second-order moments \cite{simon00,duan00,mancini02,giovannetti03,hyllus06,serafini06,fujikawa09} are adequate.  However, there is a large interest in non-Gaussian states, since non-Gaussianity is  necessary for some quantum information tasks, such as quantum computation \cite{lloyd99,bartlett02,ohliger10} and entanglement distillation \cite{dong08,hage08}. Second-order criteria are sufficient but not necessary for entanglement in non-Gaussian states.   As such, there has been some work dedicated towards entanglement detection in non-Gaussian states \cite{agarwal05, shchukin05, hillery06a, hillery06b,chen07,rodo08, walborn09,miranowicz09,hillery09,sperling09b,sperling09,adesso09}.  The set of criteria derived by Shchukin and Vogel (SV) \cite{shchukin05}, for instance, is very powerful and general,  but may require a large number of measurements \cite{shchukin05b}.  We note that the SV criteria has been applied for the experimental detection of non-Gaussian entanglement \cite{gomes09b}.  
\par
It has been shown that classical entropy functions can be used to formulate Bell's inequalities \cite{braunstein88} and entanglement witnesses for bipartite $d\times d$ level systems \cite{giovannetti04}.  These are examples of non-linear entanglement witnesses, which provide improvements in sensibility at little to no extra experimental effort \cite{guhne06b,moroder08}.      
In Ref. \cite{walborn09}, the Shannon entropy of complementary distributions was used to derive a set of entanglement witnesses for  bipartite continuous variable quantum systems. This approach is especially useful in the experimental characterization of entanglement, since it considers only a pair of joint quadrature measurements.  At the same time, these entropic witnesses are more sensitive than second-order tests (i.e  those based solely on the elements of the covariance matrix) \cite{simon00, duan00, mancini02, giovannetti03, hyllus06}. In the present work, we extend this approach by deriving entanglement criteria using more general entropy functions.   For example, we use the classical R\'enyi entropy, characterized by the continuous parameter $\alpha$, to derive a family of entropic entanglement witnesses which provides a more powerful tool for identification of entanglement.  We note that the Wehrl entropy \cite{marchiolli08}, and also quantum versions of the Shannon \cite{barnett89} and R\'enyi entropies \cite{horodecki96b,horodecki96c} have been used to identify quantum entanglement.  In general, these criteria require complete knowledge of the density matrix or more complicated measurement schemes \cite{bovino05}.     
\par
This paper is organized as follows.  In section \ref{sec:shannon} we define our notation and briefly review the criteria of Ref. \cite{walborn09}.  In section \ref{sec:improvement} we develop a family of entanglement witnesses for continuous variables using the classical R\'enyi entropy.   We then extend these results to include the more realistic case of discrete sampling.  Using these discretized distributions, we derive another set of inequalities using the discrete Tsallis entropy.  We tested the continuous variable R\'enyi criteria on several experimentally relevant states. Section \ref{sec:results} provides numerical results which show
that the generalized R\'enyi witnesses detect entanglement
in a wider variety of quantum states than second order tests or witnesses based solely on Shannon entropy \cite{walborn09}.  In section \ref{sec:conclusions} we provide concluding remarks.    
\section{Entanglement Critera with Shannon Entropy}
\label{sec:shannon}
We first review two sets of inequalities which were developed in Ref. \cite{walborn09}.  These inequalities are satisfied for all separable states, so that the violation of either one indicates that the bipartite state is entangled.

We first take into account a rotation of the usual canonical operators $\mathsf{x}$ and $\mathsf{p}$, and define a pair of general complementary operators for systems 1 and 2 as 
\begin{subequations}
\label{eq:quads}
\begin{equation}
\mathbf{\mathsf{r_j}}={\rm{cos}} {\theta_j} \mathbf{\mathsf{x_j}}+{\rm{sin}} {\theta_j} \mathbf{\mathsf{p_j}}
\end{equation}
 \begin{equation}\mathbf{\mathsf{s_j}}={\rm{cos}} {\theta_j } \mathbf{\mathsf{p_j}}-{\rm{sin}} {\theta_j} \mathbf{\mathsf{x_j}},
 \end{equation}
 \end{subequations}
 where $j=1,2$ refers to each subsystem of the bipartite state. The commutation relation $[\mathbf{\mathsf{x_j}},\mathbf{\mathsf{p_k}}]=i\delta_{j,k}$ for canonical operators $\mathbf{\mathsf{x_j}}$ and $\mathbf{\mathsf{p_k}}$ implies in $[\mathbf{\mathsf{r_j}},\mathbf{\mathsf{s_k}}]=i\delta_{j,k}$, $j,k=1,2$.  Here $x$ and $p$ are dimensionless continuous variables, such as quadratures of electromagnetic field modes or dimensionless position and momentum of a point particle, for example.   Let us define the global operators $\mathbf{\mathsf{r_{\pm}}}$ and $\mathbf{\mathsf{s_{\pm}}}$:
\begin{subequations}
    \begin{equation}
        \mathbf{\mathsf{r_{\pm}}}=\mathbf{\mathsf{r_1}} \pm \mathbf{\mathsf{r_2}},
        \label{eqGlobalOpR}
    \end{equation}
and
    \begin{equation}
        \mathbf{\mathsf{s_{\pm}}}=\mathbf{\mathsf{s_1}} \pm \mathbf{\mathsf{s_2}}.
        \label{eqGlobalOpS}
    \end{equation}
\end{subequations}
Since $[\mathbf{\mathsf{r_j}},\mathbf{\mathsf{s_k}}]=i\delta_{j,k}$, $j,k=1,2$, it is easy to see that $[\mathbf{\mathsf{r_{\mu}}},\mathbf{\mathsf{s_{\nu}}}]=2i\delta_{\mu,\nu}$ with $\mu,\nu=\pm$.

The inequalities in Ref. \cite{walborn09} were developed initially for a separable pure state $|\psi_1\rangle \otimes |\psi_2\rangle$, corresponding to the wave function $ \Psi(r_1, r_2)  = \psi_1(r_1) \psi_2(r_2)$, which can also be written as 
\begin{equation}
    \Psi(r_+, r_-)  = \frac{1}{\sqrt{2}}\psi_1\left(\frac{r_{+}+r_{-}}{2}\right) \psi_2\left(\frac{r_{+}-r_{-}}{2}\right).
\label{eqWaveFunction}
\end{equation}
For simplicity, we denote the probability distributions associated to measurement of $r_{\pm}$ as simply $R_\pm$.  They are given by
\begin{equation}
    R_{\pm}= \frac{1}{2} \int {\rm d}r_{\mp} R_{1}\left(\frac{r_{+}+r_{-}}{2}\right)R_{2}\left(\frac{r_{+}-r_{-}}{2}\right), 
    \label{eqProbDistrRpm}
\end{equation}
which is equivalent to the convolution
\begin{equation}
    R_{\pm}=R_1 * R_2^{(\pm)},
    \label{eq:Rpm}
\end{equation}
where $R_i(r_i)=|\psi_i(r_i)|^2$, $R_2^{+} \equiv R_2(r)$ and $R_2^{-} \equiv R_2(-r)$.
The Shannon entropy for continuous variables is defined by
\begin{equation}
   H[R] = -\int {\rm d}r R(r)\ln R(r),
\label{eqDefShannonCont}
\end{equation} where $R(r)$ is the probability distribution associated to the measurement of an arbitrary continuous variable $r$. Similar expressions are obtained for the probability distribution $S$ of the complementary variable $s$.

Two inequalities were introduced in Ref. \cite{walborn09}. Their violation indicates the presence of entanglement.  Using the probability distributions $R_\pm$ and $S_\pm$ defined above and applying the entropy power inequality \cite{cover}, the following criteria were obtained:
\begin{eqnarray}
    H[R_{\pm}]+ H[S_{\mp}] \geq \frac{1}{2}\ln \left[ \sum_{i,j} e^{(2 H[R_i]+2 H[S_j])} \right].
\label{eqStrong}
\end{eqnarray}
These criteria are useful only in the case of pure states.  They can be extended to include mixed states as well, but numerical optimization procedures are required \cite{walborn09}.   By further applying an entropic uncertainy relation for the distributions $R_j$ and $S_j$ \cite{bialynicki75},  
a second set of entropic witnesses were derived:
\begin{equation}
    H[R_{\pm}]+ H[S_{\mp}] \geq \ln (2\pi e).
\label{eqWeak}
\end{equation}
Although inequality \eqref{eqWeak} is weaker than inequality \eqref{eqStrong}, it has the advantage that it is also suitable for mixed states.  It was shown that both of these criteria are more sensitive than second-order tests involving the same operators.
\section{GENERALIZATION OF ENTROPIC CRITERIA}
\label{sec:improvement}
A natural attempt to improve the entropic entanglement witnesses described in Section II is the application of a more general function of information entropy. For this purpose, we first employ the R\'{e}nyi  entropy for continuous variables, defined by \cite{renyi61,cover}
\begin{equation}
    H_{\alpha} [R] = \frac{1}{1-\alpha}\ln \left[\int {\rm d}r R^\alpha (r) \right ]=\frac{\alpha}{1-\alpha}\ln \|R\|_\alpha,
    \label{eqDefRenyiCont}
\end{equation}where ${\|R\|}_{\alpha}$ is the $\mathcal{L}_{\alpha}$ norm of the distribution $R$ (see Ref. \cite{cover}):

\begin{equation}
        \|R\|_\alpha  = \left[ \int {\rm d}r R^\alpha (r)\right ]^{1/\alpha}.
\label{eqDefNorm}
\end{equation}

As in section \ref{sec:shannon}, let us first consider only pure states of the form $|\psi_1\rangle \otimes |\psi_2\rangle$. Using the probability distributions \eqref{eq:Rpm}, the R\'{e}nyi entropy for global distributions is
\begin{equation}
    H_{\alpha} [R_{\pm}] = \frac{\alpha}{1-\alpha}\ln \|R_1 * R_2^{(\pm)}\|_\alpha.
    \label{eqRenyiGlobal}
\end{equation}
To derive an inequality, we employ Young's inequality, which is valid for convolutions of distributions \cite{barthe98}. For $1/\alpha = 1/{\alpha_1}+1/{\alpha_2}-1$, Young's inequality is 
\begin{equation}
    \|R_1 * R_2^{(\pm)}\|_\alpha \leq C(\alpha_1, \alpha_2) \|R_1\|_{\alpha_1}  \|R_2\|_{\alpha_2},
    \label{eqYoung1}
\end{equation} 
with $\alpha, \alpha_1, \alpha_2 \geq1$ or
\begin{equation}
    \|R_1 * R_2^{(\pm)}\|_\alpha \geq C(\alpha_1, \alpha_2) \|R_1\|_{\alpha_1}  \|R_2\|_{\alpha_2},
\label{eqYoung2}
\end{equation}
for $\alpha, \alpha_1, \alpha_2 \leq1$.  The coefficent $C(\alpha_1, \alpha_2) $ is given by
\begin{equation}
C(\alpha_1, \alpha_2) = \frac{C_{\alpha_1} C_{\alpha_2}}{C_{\alpha}},
\label{eqYoungCoeff}
\end{equation}
where 
\begin{equation}
C_t= \sqrt{\frac{t^{\frac{1}{t}}}{|t'|^{\frac{1}{t'}}}},
\end{equation}
with $t^\prime \equiv t/(t-1)$.
Without loss of generality, we choose variables  such that $\alpha$, $\alpha_1$, $\alpha_2 \geq1$ and $0 \leq \beta$, $\beta_1$, $\beta_2 \leq1$. Then, from inequalities \eqref{eqYoung1} and \eqref{eqYoung2} we can write: 
\begin{subequations}
    \begin{equation}
        \|R_{\pm}\|_\alpha \leq C(\alpha_1, \alpha_2) \|R_1\|_{\alpha_1}  \|R_2\|_{\alpha_2},
        \label{eqYoung3}
    \end{equation}
and
    \begin{equation}
        \|S_{\mp}\|_\beta \geq C(\beta_1, \beta_2) \|S_1\|_{\beta_1}  \|S_2\|_{\beta_2},
    \label{eqYoung4}
    \end{equation}
\end{subequations}
where we remember that 
\begin{subequations}
\label{eqRelations}
    \begin{eqnarray}
        \frac{1}{\alpha} = \frac{1}{\alpha_1} + \frac{1}{\alpha_2} -1,
        \label{eqRelationsAlphas}
    \end{eqnarray}
and
    \begin{eqnarray}
        \frac{1}{\beta} = \frac{1}{\beta_1} + \frac{1}{\beta_2} -1.
        \label{eqRelationsBetas}
    \end{eqnarray}
\end{subequations}
Dividing inequality \eqref{eqYoung3} by inequality \eqref{eqYoung4}, we can set up a new inequality
\begin{equation}
    \frac{\|R_{\pm}\|_\alpha}{\|S_{\mp}\|_\beta} \leq \frac{C(\alpha_1, \alpha_2)}{C(\beta_1, \beta_2)} \frac{\|R_1\|_{\alpha_1}}{\|S_1\|_{\beta_1}}  \frac{\|R_2\|_{\alpha_2}}{\|S_2\|_{\beta_2}},
    \label{eqYoungDiv}
\end{equation} 
which will be verified when the pure state is separable,  since the distributions $R_{\pm}$ and $S_{\mp}$ can be expressed in terms of convolutions of the probability distributions of the two subsystems. 
\par
We can write the norm in terms of an entropy, such as the R\'{e}nyi entropy or Tsallis entropy.  Taking the logarithm of inequality (\ref{eqYoungDiv}) and using Eq. \eqref{eqDefRenyiCont} results in an inequality in terms of R\'{e}nyi entropies:

\begin{align}
    &\left(\frac{\alpha-1}{\alpha} \right) H_\alpha[R_{\pm}]+
    \left( \frac{1-\beta}{\beta} \right) H_\beta[S_{\mp}] \geq \nonumber\\
   &\left( \frac{\alpha_1-1}{\alpha_1} \right) H_{\alpha_1} [R_1] +
    \left( \frac{1-\beta_1}{\beta_1} \right) H_{\beta_1} [S_1] +\nonumber\\
    &\left( \frac{\alpha_2-1}{\alpha_2} \right) H_{\alpha_2} [R_2] +
    \left( \frac{1-\beta_2}{\beta_2} \right) H_{\beta_2} [S_2] + \nonumber \\
     &\ln \left[\frac{C(\beta_1, \beta_2)}{C(\alpha_1, \alpha_2)}\right]\ .
    \label{eqStrongRenyi}
\end{align}
Inequality \eqref{eqStrongRenyi} is a generalization of criteria \eqref{eqStrong}. In order to recover \eqref{eqStrong} from \eqref{eqStrongRenyi} we first consider the case $\alpha=\beta$
and then take the limit $\alpha \rightarrow 1$.  Violation of inequality \eqref{eqStrongRenyi} implies that the pure state considered is  entangled.  Extension of \eqref{eqStrongRenyi} to include mixed states is possible, although evaluation of the right-hand side requires minimization over all possible decompositions of the mixed state, and as such, is not very useful in an experimental setting \cite{walborn09}.     
\par
To derive a second inequality that does not depend on the entropy functions $H_{\alpha_j}[R_j]$ and $H_{\beta_j}[S_j]$,  we employ the entropic uncertainty relation for R\'{e}nyi entropy given by Ref. \cite{bialynicki06}:
\begin{equation}
    H_{\alpha_j}[R_j]+ H_{\beta_j}[S_j] \geq
    -\frac{1}{2(1-\alpha_j)} \ln \frac{\alpha_j}{\pi}-\frac{1}{2(1-\beta_j)} \ln \frac{\beta_j}{\pi},  
    \label{eqBirula38}
\end{equation}
where it is necessary to include the restriction \cite{bialynicki06}:
\begin{equation}
    \frac{1}{\alpha_j}+\frac{1}{\beta_j}=2,\,\, j=1, 2.
    \label{eqAlphaBeta12}
\end{equation}
Eq. (\ref{eqAlphaBeta12}), along with Eqs. (\ref{eqRelations}), lead to
\begin{equation}
    \frac{1}{\alpha}+\frac{1}{\beta}=2.
    \label{eqAlphaBeta}
\end{equation}
Applying the uncertainty relation (\ref{eqBirula38}) to inequality (\ref{eqStrongRenyi}) and performing  some algebra we obtain the inequality 
\begin{align}
    &H_\alpha[R_{\pm}]+ H_\beta[S_{\mp}] \geq \nonumber \\ &-\frac{1}{2(1-\alpha)} \ln \frac{\alpha}{\pi}-\frac{1}{2(1-\beta)} \ln \frac{\beta}{\pi} +\nonumber \\ &
   \frac{\alpha}{\alpha-1} \sum\limits_{j=1,2} \frac{\alpha_j-1}{\alpha_j}\ln \left|\frac{\alpha_j}{\alpha_j-1}\right | -\ln \left|\frac{\alpha}{\alpha-1} \right|.
    \label{eqBirulaInt}
\end{align}
The sum of terms in the last line of Eq. \eqref{eqBirulaInt} is always non-negative.  $\alpha_1$ and $\alpha_2$ are arbitrary parameters within the restrictions imposed by Eqs. \eqref{eqRelationsAlphas} and \eqref{eqAlphaBeta}, which guarantee that $1 \leq 1/\alpha_1+1/\alpha_2 \leq 2$.   Within this domain we can maximize the last term on the right-hand side of inequality \eqref{eqBirulaInt}, which reaches a maximum value of $\ln 2$ when $\alpha_1=\alpha_2$.   This leads directly to the inequality:
\begin{align}
    H_\alpha[R_{\pm}]+ H_\beta[S_{\mp}] \geq-\frac{1}{2(1-\alpha)} \ln \frac{\alpha}{2 \pi}-\frac{1}{2(1-\beta)} \ln \frac{\beta}{2 \pi}.
    \label{eqBirulaWeak}
\end{align}
Note that our choice $\alpha \geq 1$ and $1/2 \leq \beta \leq 1$ is arbitrary, and that these restrictions can be switched with no alteration in the derivation.  Inequality \eqref{eqBirulaWeak} reduces to \eqref{eqWeak} when $\alpha \longrightarrow 1$.  
\par
We'll now show that inequality (\ref{eqBirulaWeak}) is also valid for mixed states.  Noting that $[\mathbf{\mathsf{r_{\mu}}},\mathbf{\mathsf{s_{\nu}}}]=2i\delta_{\mu,\nu}$, ($\mu,\nu=\pm$), then the uncertainty relation for the R\'enyi entropy of complementary distributions $R_\pm$ and $S_\pm$ is 
   \begin{equation}
    H_{\alpha}[R_\pm]+ H_{\beta}[S_\pm] \geq
    -\frac{1}{2(1-\alpha)} \ln \frac{\alpha}{2\pi}-\frac{1}{2(1-\beta)} \ln \frac{\beta}{2\pi},  
    \label{eqBirulapm}
\end{equation}
where again $1/\alpha + 1/\beta = 2$.  Bialynicki-Birula has shown that this uncertainty relation is also valid for mixed states \cite{bialynicki06}, in which case $R_\pm$ and $S_\pm$ are complementary marginal distributions obtained from the Wigner function associated to the mixed quantum state.   
We can now make use of an alternative way of deriving inequality (\ref{eqBirulaWeak}) by means of the positive partial transpose (PPT) criterion \cite{peres96,horodecki96,simon00}.   For any continuous variable quantum state, the transpose operation is equivalent to a mirror reflection in phase space, taking $(r_j,s_j)\longrightarrow (r_j,-s_j)$ \cite{simon00}.  Thus, the partial transpose of a bipartite state $\varrho_{12}$ thus takes the global variables $r_\pm \longrightarrow r_\pm$ and $s_\pm \longrightarrow s_\mp$, where we take the transpose of subsystem 2.   The marginal probability distributions under partial transposition $T$  transform as 
\begin{subequations}
\label{eq:pt2}
\begin{align}
R_{\pm}^T & =R_{\pm}  \\
S^T_{\pm}& =S_{\mp},
\end{align}
\end{subequations}
and we have
  \begin{equation}
    H_{\alpha}[R^T_\pm]+ H_{\beta}[S^T_\pm] = H_{\alpha}[R_\pm]+ H_{\beta}[S_\mp]. 
       \label{eq:pt}
\end{equation}
The partial transpose of a separable density operator is a positive operator, and thus it is still a physical state \cite{peres96, horodecki96, simon00}, and will satisfy the uncertainty relation
\eqref{eqBirulapm}.  
 Substituting Eq. \eqref{eq:pt} into inequality \eqref{eqBirulapm} leads directly to inequality (\ref{eqBirulaWeak}), where we have made no assumptions about the purity of the bipartite state $\varrho_{12}$.  Thus,  criteria (\ref{eqBirulaWeak}) is also valid for bipartite mixed states.  
 \par
 The above argument illustrates that the family of entropic entanglement witnesses \eqref{eqBirulaWeak} are in fact PPT criteria.  This illustrates a general method for developing new PPT criteria:  apply \emph{any} quantum mechanical uncertainty relation to distributions $R_\pm$ and $S_\pm$, and use Eqs. \eqref{eq:pt2}.  We note that this was the general spirit of the procedure used by Simon to develop a criteria based on second-order moments \cite{simon00}, and has also been used in Ref. \cite{nha08}.    
\subsection{Relationship with second-order criteria}
The second-order Mancini-Giovannetti-Vitali-Tombesi (MGVT) criteria is \cite{mancini02}
\begin{equation}
\Delta_{r_{\pm}}^2 \Delta_{s_{\mp}}^2 \geq 1,   
\label{eq:MGVT}
\end{equation}
where $\Delta_{q}^2$ is the variance in variable $q$.  Inequality \eqref{eq:MGVT} is verified by any separable state.      
In Ref. \cite{walborn09}, it was shown that the MGVT criteria can be derived directly from the Shannon criteria \eqref{eqWeak} by maximizing the sum $H[R_\pm]+H[S_\mp]$.  This leads to the inequalities:
\begin{equation}
   \ln (2 \pi e \Delta_{r_{\pm}} \Delta_{s_{\mp}}) \geq H[R_{\pm}]+ H[S_{\mp}] \geq \ln (2\pi e).
\label{eq:limit}
\end{equation}
This upper bound is saturated for Gaussian probability distributions \cite{shannon}.  Since $R_\pm$ and $S_\pm$ are arbitrary (though complementary) marginal distributions in phase space, this implies that the bound is saturated for Gaussian states.  Nevertheless,
within the class of  non-Gaussian states, inequalities \eqref{eq:limit} 
show that the criteria given in \eqref{eqWeak} may detect entanglement in states that the MGVT criterion might not \eqref{eq:MGVT}.   
\par A natural question to ask is whether we can derive new entanglement witnesses by maximizing the sum of R\'enyi entropies $H_\alpha[R_\pm]+H_\beta[S_\mp]$ in criteria \eqref{eqBirulaWeak}.  Doing so leads to an inequality also involving second-order moments, due to the fact that the R\'enyi entropy is maximized for the Student-t  and Student-r distributions \cite{costa02,vignat06}, which (for zero mean) are completely characterized  by the variance.  More specifically, we arrive at    
\begin{equation}
\Delta_{r_{\pm}}^2 \Delta_{s_{\mp}}^2 \geq f(\alpha,\beta),  
\label{eq:Toscano}
\end{equation}     
where $f(\alpha,\beta) \leq 1$ for all allowed values of $\alpha$
and $\beta$.  In the limiting case $\alpha, \beta \longrightarrow 1$,  $f(\alpha,\beta) = 1$ and we recover the MGVT criteria \eqref{eq:MGVT}.  Thus, inequality \eqref{eq:Toscano} is not an improvement over the already established MGVT criterion.  
\section{Discrete Distributions}
\subsection{Discrete R\'enyi Entropy}
Inequalities \eqref{eqStrongRenyi} and \eqref{eqBirulaWeak} derived in the above section were developed for continuous distributions $R_\pm$ and $S_\pm$.  However, in an experimental setting one typically measures discrete distributions, due to the finite resolution of the measurement apparatus.  Here, we show how to deal with discrete resolution and we derive an entanglement witness
equivalent to \eqref{eqBirulaWeak}, but for discrete distributions. The same procedure can be adopted
for a derivation of inequalities  equivalent to \eqref{eqStrongRenyi}. Let us call these discrete distributions $R^\delta_\pm$ and $S^\Delta_\pm$, and suppose that their elements are    
\begin{subequations}
\label{eq:discretedist}
\begin{equation}
\rho^{\delta}_{k \pm} = \int\limits_{k \delta}^{(k+1)\delta} R_{\pm}(r) dr  
\end{equation}
and
\begin{equation}
\sigma^{\Delta}_{k \pm} = \int\limits_{k \Delta}^{(k+1)\Delta} S_{\pm}(s) ds,  
\end{equation}
\end{subequations}
respectively.  Here we assume that $r$ measurements have resolution $\delta$ and $s$ measurements are performed with resolution $\Delta$.  To apply these inequalities to discrete distributions, one can write the entropy of the continuous distribution 
in terms of the discrete distribution as \cite{cover}
\begin{subequations}
\begin{align}
H_\alpha[R_\pm] & = H_\alpha[R^\delta_\pm] + \ln \delta, \\
H_\beta[S_\pm] & = H_\beta[S^\Delta_\pm] + \ln \Delta,
\end{align}
\end{subequations}
provided that $\delta$ and $\Delta$ are sufficiently small.  
Here the discrete R\'enyi entropy is 
\begin{equation}
H_\alpha[R^\delta_\pm]= \frac{1}{1-\alpha} \ln \left ( \sum\limits_k \left (\rho^\delta_{k \pm}\right )^\alpha \right) , 
\end{equation}
and similarly for $H_\beta[S^\Delta_\pm]$.
Inequality \eqref{eqBirulaWeak} can then be written in terms of the discrete distributions:
\begin{align}
    H_\alpha[R^\delta_{\pm}]+ H_\beta[S^\Delta_{\mp}] \geq-\frac{1}{2}\left(\frac{\ln\alpha}{1-\alpha}+\frac{\ln\beta}{1-\beta}\right)+\ln\left(\frac{2\pi}{\delta\Delta}\right).
    \label{eqBirulaWeakDisc}
\end{align}
Nevertheless, the above inequalities are also valid for arbitrary size of the resolutions $\delta$ and $\Delta$, since it is also possible to derive them by direct application of the uncertainty relation for the discrete R\'enyi entropies, as developed by Bialynicki-Birula \cite{bialynicki06}.    

\subsection{Entanglement Criteria with Tsallis Entropy}
An uncertainty relation for the Tsallis entropy \cite{tsallis} $T_\alpha$ of continuous distributions was developed in Ref. \cite{rajagopal95}.  Since the R\'enyi entropy of a continuous distribution can be written as $H_\alpha = \ln[1+(1-\alpha)T_\alpha]/(1-\alpha)$, one can show that this uncertainty relation is equivalent to that of Ref. \cite{bialynicki06}, given in Eq. \eqref{eqBirula38}.  Thus, development of entanglement witnesses based on the uncertainty relation in Ref. \cite{rajagopal95} would be equivalent to those developed in the previous section.  Recently, Wilk and Wlodarczyk (WW) \cite{wilk08} have derived an uncertainty relation for the discrete Tsallis entropy that is distinct from the discretization of the relation given in Ref. \cite{rajagopal95}.  Furthermore, the WW relation cannot be extended to include continuous distributions.  We will thus employ the WW relation to arrive at a new set of entanglement criteria based on discrete Tsallis entropy.       
\par     
The Tsallis entropy for a discrete random variable $X$ is defined as \cite{tsallis} 
\begin{equation}
T_\alpha[X] = \frac{1}{1-\alpha}\left (\sum_k x_k^\alpha -1 \right).   
\end{equation}
The WW uncertainty relations for discrete distributions $R^\delta_\pm$ and $S^\Delta_\pm$ are, in the case $\delta \Delta \leq (2\pi/\beta)(\alpha/\beta)^{1/2(\alpha-1)}$ \cite{wilk08}, 
\begin{equation}
T_{\alpha}[R^\delta_\pm]+T_{\beta}[S^\Delta_\pm] \geq \frac{1}{1-\alpha}\left [\left(\frac{\beta}{\alpha}\right)^{1/2\alpha}\left(\frac{\beta \delta \Delta}{2 \pi} \right)^{(\alpha-1)/\alpha} -1 \right]
\label{eq:WW1}
\end{equation}
and,  in the case $\delta \Delta > (2 \pi/\beta)(\alpha/\beta)^{1/2(\alpha-1)}$, 
\begin{equation}
T_{\alpha}[R^\delta_\pm]+T_{\beta}[S^\Delta_\pm] \geq \frac{1}{\alpha-1}\left [\left(\frac{\alpha}{\beta}\right)^{1/2\alpha}\left(\frac{\beta \delta \Delta}{2 \pi} \right)^{(1-\alpha)/\alpha} -1 \right].  
\label{eq:WW2}
\end{equation}
It follows directly from Eqs. \eqref{eq:discretedist} that the partial transpose takes $(R^\delta_\pm)^{T}\longrightarrow R^\delta_\pm$ and $(S^\Delta_\pm)^{T}\longrightarrow S^\Delta_\mp$, and we have  
\begin{equation}
T_{\alpha}[(R^\delta_\pm)^{T}]+T_{\beta}[(S^\Delta_\pm)^{T}]=T_{\alpha}[R^\delta_\pm]+T_{\beta}[S^\Delta_\mp].
\label{eq:Tpt}
\end{equation}
Since any separable state should still be a physical state under partial transpose, the distributions $(R^\delta_\pm)^{T}$ and $(S^\Delta_\pm)^{T}$ of any separable state satisfies inequalities \eqref{eq:WW1} and \eqref{eq:WW2}.  Using Eq. \eqref{eq:Tpt} in uncertainty relations \eqref{eq:WW1} and \eqref{eq:WW2} leads immediately to entanglement witnesses using the Tsallis entropy.  Violation of either inequality thus implies implies that the quantum state associated
with the discrete marginal distributions 
$R_{\pm}^{\delta}$ and 
$S_{\pm}^{\Delta}$ is entangled.   
\section{Examples}
\label{sec:results}
Here we provide some examples which show the utility of the R\'enyi entropic criteria presented in section \ref{sec:improvement}.  We focus on several examples of continuous variable states which are currently of experimental interest.  We leave further numerical investigation to future work.    
\subsection{Hermite-Gauss state}
Consider the non-Gaussian state given by  
\begin{eqnarray}
    \eta (r_1, r_2) = \frac{(r_1+r_2)}{\sqrt{\pi \sigma_{-} \sigma_{+}^3}} e^{-(r_1+r_2)^2/4\sigma_{+}^2} e^{-(r_1-r_2)^2/4\sigma_{-}^2}, \nonumber\\
    \label{eqNonGaussianState}
\end{eqnarray}
where the widths
$\sigma_{+}$ and $\sigma_{-}$ characterize the state.  State \eqref{eqNonGaussianState} is non-separable for any value of parameters $\sigma_{+}$ and $\sigma_{-}$.  This state has been experimentally produced using spontaneous parametric down-conversion and been shown to have several interesting properties \cite{walborn03b,nogueira04a,gomes09a,gomes09b}.  We note that it is equivalent to the single-photon entangled state considered in Ref. \cite{agarwal05},  when $\sigma_+=\sigma_-=1$.  
\par
\begin{figure}
\begin{center}
\includegraphics[width=7cm]{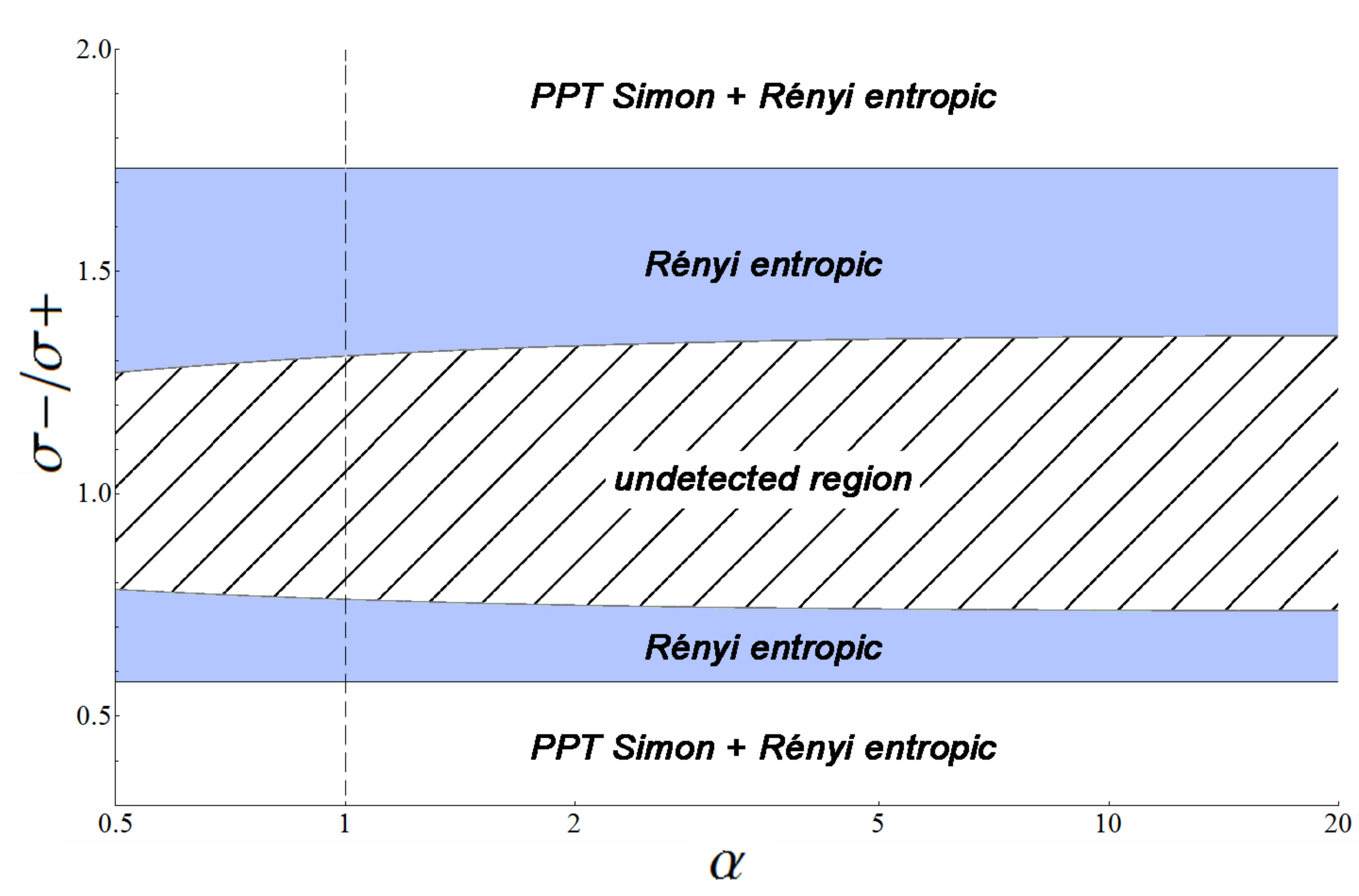}
\end{center}
\caption{\footnotesize  (color online) Entanglement detection of state \eqref{eqNonGaussianState}.  The light blue shaded region is where the R\'{e}nyi entropic criteria in (\ref{eqBirulaWeak}) identifies entanglement, while the Simon second-order PPT criterion does not.  The uppermost and lowermost areas designate the regions in which the Simon PPT and R\'enyi criteria detect entanglement in state \eqref{eqNonGaussianState}.  In the center hatched region neither test detect entanglement.}
\label{comparingnongaussian}
\end{figure}

The application of the witness \eqref{eqBirulaWeak}, after a lengthy but straightforward calculation, leads to:
\begin{subequations}
\begin{equation}
   \frac{\sigma_{-}}{\sigma_{+}} <
   \left[\frac{\pi^{\frac{1}{2}}}{\Gamma(\alpha+\frac{1}{2})}\left(\frac{\alpha}{2}\right)^\alpha\right]^\frac{1}{1-\alpha},
\label{eqsiginf}
\end{equation}

\begin{equation}
   \frac{\sigma_{-}}{\sigma_{+}} >
   \left[\frac{\pi^{\frac{1}{2}}}{\Gamma(\alpha+\frac{1}{2})}\left(\frac{\alpha}{2}\right)^\alpha\right]^{-\frac{1}{1-\alpha}}, 
\label{eqsigsup}
\end{equation}
\end{subequations}
where we have included both cases:  $\alpha \geq 1$ and $1/2 \leq \alpha \leq 1$.  
Thus, only entangled states of the form \eqref{eqNonGaussianState} that violate one of these inequalities are detected by our entropic R\'enyi criteria \eqref{eqBirulaWeak}.  For $\alpha=1$ the limits $\sigma_{-} / \sigma_{+} < \frac{e^{1-\gamma}}{2}$ and $\sigma_{-} / \sigma_{+} > \frac{2}{e^{1-\gamma}}$ ($\gamma$ is the Euler's constant)  obtained in \cite{walborn09} are recovered.
Figure \ref{comparingnongaussian} shows the limits of entanglement detection of the state \eqref{eqNonGaussianState} as a function of $\alpha$. The graph shows that we improve sensibility using the R\'{e}nyi entropic inequality \eqref{eqBirulaWeak} when $\alpha \longrightarrow 1/2$. For example, in the particular case of the non-Gaussian state $\eta(r_1,r_2)$ with $\sigma_{-} / \sigma_{+} = 1.3$, entanglement is not detected by the Shannon entropy criterion of (\ref{eqWeak}), but it is detected by the more general R\'enyi entropy criterion (\ref{eqBirulaWeak}).  At the same time, there is a large region ($1/\sqrt{3} < \sigma_-/\sigma_+ < \sqrt{3}$) where the second-order Simon criterion \cite{simon00} does not detect entanglement in state \eqref{eqNonGaussianState}.   The Simon criterion is a necessary and 
sufficient condition for entanglement in bipartite Gaussian states. So, in the case where 
the Simon criterion fails to detect entanglement, the covariance matrix of the state is ``separable",
or in other words,  the bipartite Gaussian state with the same covariance matrix is separable.  Thus, we can guarantee that any second-order entanglement criterion also fails to detect entanglement in this region.
\subsection{NOON States}
\begin{figure}
\begin{center}
\includegraphics[width=8cm]{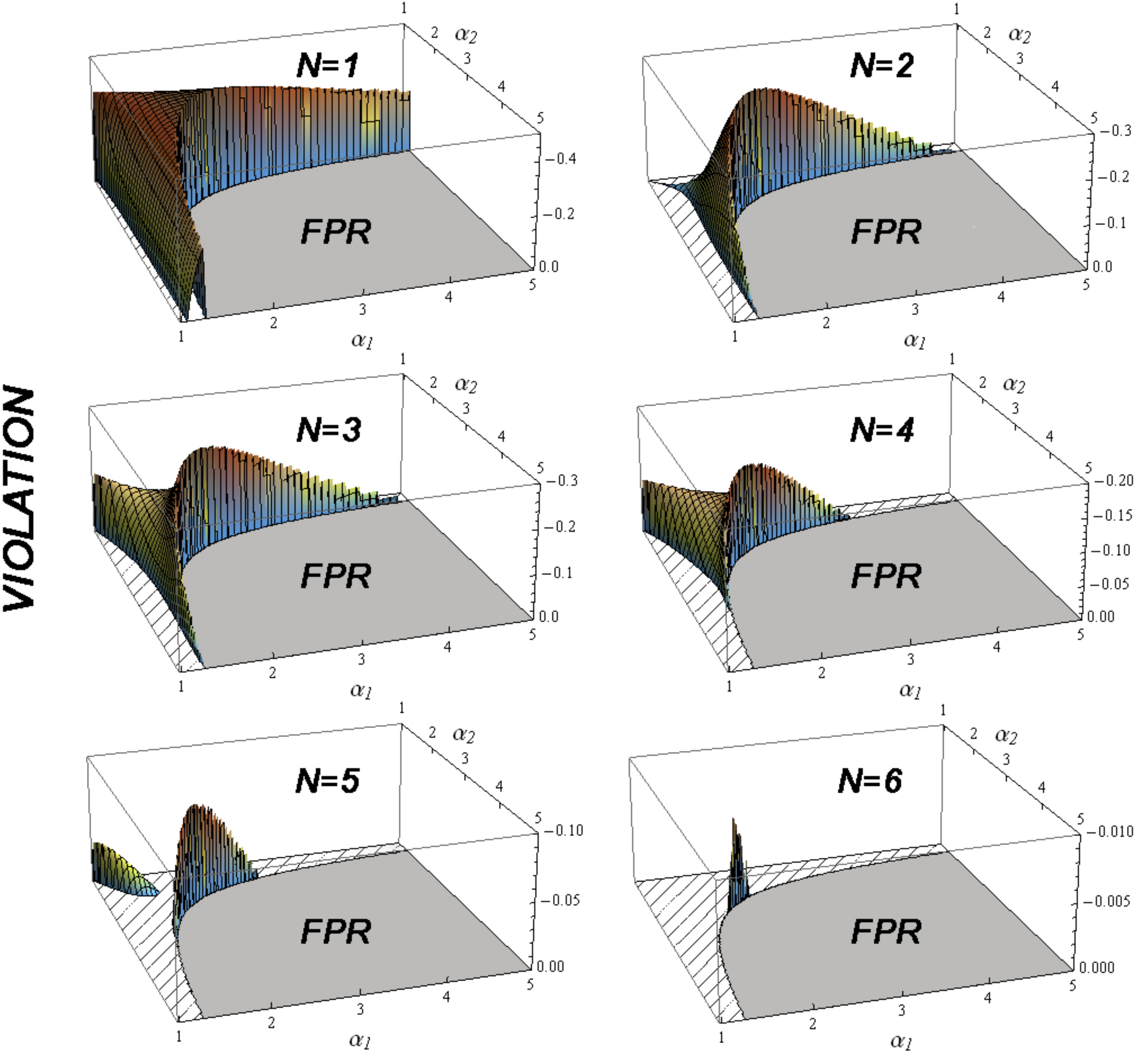}
\end{center}
\caption{\footnotesize  (color online) Entanglement detection for NOON state for $N=1$ to $6$. The surfaces represents the regions where the strong R\'{e}nyi entropic criteria \eqref{eqStrongRenyi} detects entanglement as function of $\alpha_1$ and $\alpha_2$.  The criteria were tested for $\theta_j=0$.  FPR designates the ``forbidden parameter region", as determined by Eqs. \eqref{eqRelationsAlphas} and \eqref{eqRelationsBetas}.}
\label{comparingNOON}
\end{figure}

There is a lot of interest in generating entangled ``NOON" states of the form
\begin{equation}
\ket{\psi}_{NOON} = \frac{1}{\sqrt{2}} (\ket{N}_1\ket{0}_2 + \ket{0}_1\ket{N}_2), 
\end{equation}
where $\ket{n}$ is an $n$-photon Fock state.  NOON state are 
 particularly useful for quantum metrology \cite{mitchell04}.  Here we consider detection of entanglement using continuous variable quadrature measurements.    
For NOON states the inequality (\ref{eqBirulaWeak}) does not detect entanglement for any value of $\alpha$ (tested for $N \leq 10$). However, we have investigated their entanglement detection with the stronger R\'{e}nyi criterion \eqref{eqStrongRenyi}. The results are shown in Figure \ref{comparingNOON}.  We have studied the violation of inequality (\ref{eqStrongRenyi}) as a function of parameters $\alpha_1$, $\alpha_2$,  $\beta_1$, $\beta_2$. In order to simplify the calculations, we have constrained $\beta_1$ and $\beta_2$ as functions of $\alpha_1$ and $\alpha_2$, according to restriction (\ref{eqAlphaBeta12}) (see Figure \ref{comparingNOON}).  The best violations were found for $\alpha_1=\alpha_2=2$.  In all cases, we chose quadrature operators \eqref{eq:quads} with $\theta=0$.  With this choice of parameters we were able to detect entanglement up to $N=6$, which is an improvement over the Shannon criteria (\ref{eqStrong}) \cite{walborn09}.  Numerical results show that entanglement in the NOON states goes undetected under any second-order criteria (tested for $N \leq 10$).   

\subsection{Dephased Cat State}

\begin{figure}
\begin{center}
\includegraphics[width=8cm]{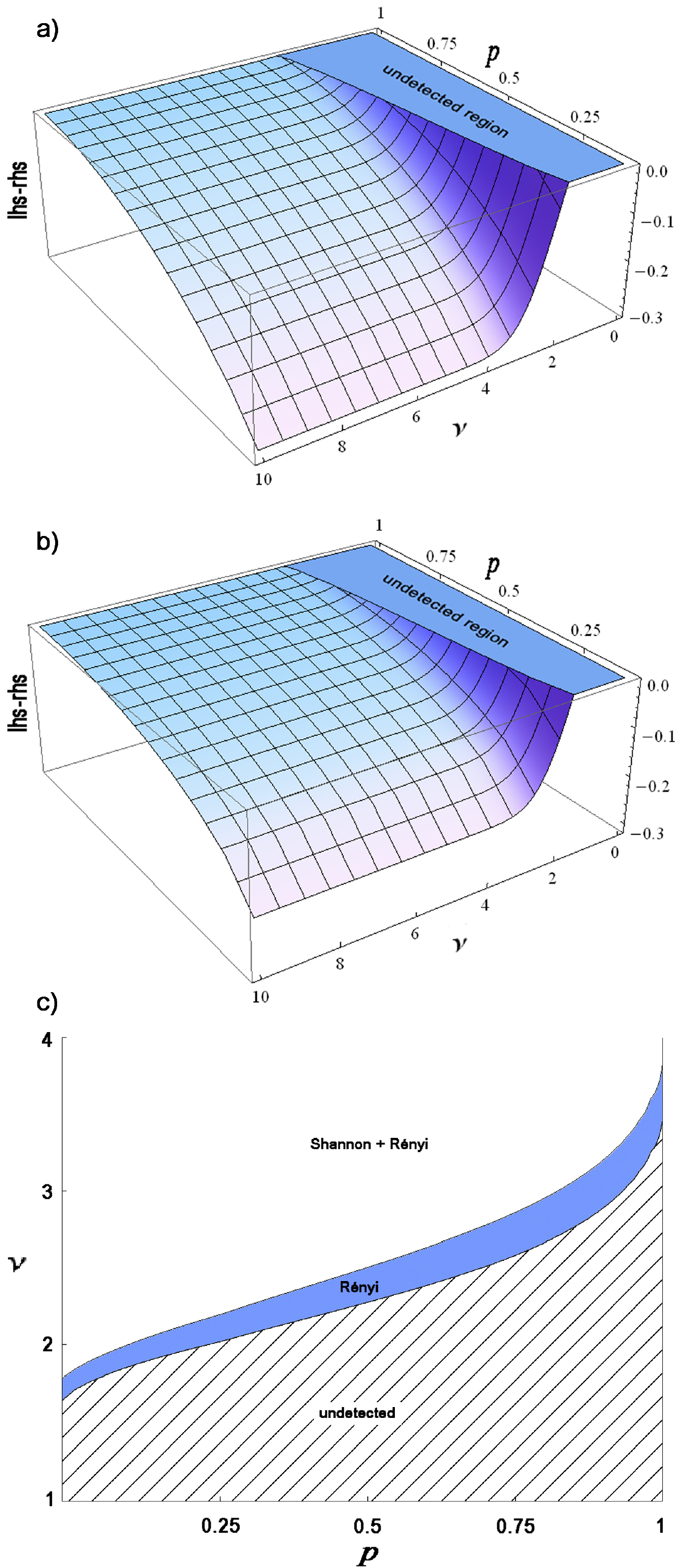}
\caption{\footnotesize  (color online) a) Violation of Shannon entanglement criterion, given by the difference of the left-hand side (lhs) and right-hand side (rhs) of (\ref{eqWeak}) for the dephased cat state (\ref{eqCatState}). (b) violation of R\'{e}nyi entanglement criterion ($\alpha$ very close to $1/2$), given by the difference of the left-hand side (lhs) and right-hand side (rhs) of (\ref{eqBirulaWeak}) for the dephased cat state (\ref{eqCatState}). c) Comparison of Shannon criteria and R\'enyi criteria.  The white region is detected by Shannon and R\'{e}nyi entropic criteria, the blue one is detected only by R\'{e}nyi entropic criterion ($\alpha\longrightarrow 1/2$) and the hatched area represents the region which remains undetected as function of $\nu$ and $p$.}
\label{cat}
\end{center}
\end{figure}

Entangled Schr\"odinger cat states have been produced experimentally in quadrature variables of two single mode fields using optical parametric amplification \cite{ourjoumtsev09}. Due to experimental imperfections these states are mixed. Here we consider mixed states given 
by the dephased entangled cat states, 
\begin{eqnarray}
    \rho=N(\nu,p) \{ |\nu, \nu \rangle \langle \nu, \nu |+|-\nu, -\nu \rangle \langle -\nu,-\nu|\nonumber\\
    -(1-p)(|\nu, \nu \rangle \langle -\nu, -\nu |+|-\nu, -\nu \rangle \langle \nu, \nu |) \}, \label{eqCatState}
\end{eqnarray}
where $N(\nu,p)$ is a normalization constant.   Parameter $p$ characterizes the dephasing \cite{chuang00}, and $\nu$ is the complex amplitude of the coherent state $|\nu\rangle$. Ref. \cite{walborn09} showed that the Shannon criteria \eqref{eqWeak} identifies entanglement for a broad range of values of parameters $p$ and $\nu$ \cite{walborn09}, which we reproduce in Figure  \ref{cat} a) for comparison. The R\'{e}nyi entropic criterion (\ref{eqBirulaWeak}) with $\alpha$ very close to  $1/2$ extends entanglement detection, as we can see in Figure \ref{cat} b).  Figure \ref{cat} c) compares these two results.  

\section{Conclusions}
\label{sec:conclusions}
We have presented a family of entanglement witnesses using generalized classical entropy 
functions applied to marginal probability distributions $R_{\pm}$ and $S_{\pm}$ 
associated with the measurement of global canonical operators $r_{\pm}$ and 
$s_{\pm}$ in continuous variable systems. First, we employed the R\'enyi entropy 
(parameterized by $\alpha$) for continuous distributions to arrive at a set of inequalities 
(see Eq. \eqref{eqStrongRenyi}) which are satisfied for all pure bipartite separable states.
Second, we introduced a set of inequalities in Eq.\eqref{eqBirulaWeak}, using also the R\'enyi entropy of 
continuous distributions, which are satisfied for all bipartite states (pure or mixed). 
We have demonstrated that these criteria offer a greater sensitivity to detection of entanglement.  We illustrated this point with several examples where the R\'enyi entropic criteria 
identify entanglement, while the Shannon entropic criteria \cite{walborn09} and second-order criteria do not \cite{simon00}.  We also showed that the entropic 
criteria given in Eq.\eqref{eqBirulaWeak}  are in fact PPT criteria, and gave a general recipe to obtain new PPT criteria based on marginal probability distributions 
$R_{\pm}$ and $S_{\pm}$ in continuous variable systems.
\par
The entanglement witnesses presented here should be very convenient in an experimental setting, as they involve a relatively small number
of measurements. In particular, fixing the local rotations involved in the definition of the global operators
$r_{\pm}$ and $s_{\pm}$, it is necessary to determine only the probability distributions $R_{\pm}$ and $S_{\pm}$.
This can be done directly via measurement of $r_{\pm}$ and $s_{\pm}$ or from measurement 
of the joint probability distributions $R(r_1,r_2)$ and $S(s_1,s_2)$.
In order to take into account the precision of the measurement apparatus we extended our R\'enyi entropy criteria \eqref{eqBirulaWeak} to include
discrete distributions (see Eq. \eqref{eqBirulaWeakDisc}).
In this case,  we also developed an entropic entanglement criteria based on the Tsallis entropy.

In addition to practical relevance,  the improvement offered by the entropic entanglement criteria is interesting from a theoretical point of view, since there is an entire family of entropic
inequalities parameterized by the order of the R\'enyi entropy (a
continuous quantity) that could be explored.
Moreover, these results encourage the use of other types
of entropy functionals and/or uncertainty relations for entanglement characterization.

\begin{acknowledgments}
We would like to thank M. O. Hor-Meyll for useful discussions, and acknowledge financial support from the Brazilian funding agencies CNPq and FAPERJ.  This work was performed as part of the Brazilian Instituto Nacional de Ci\^{e}ncia e Tecnologia - Informa\c{c}\~{a}o Qu\^{a}ntica (INCT-IQ).
\end{acknowledgments}

\end{document}